# Integrated elliptical III-V nanopillar arrays on silicon for dual photodetection and light emission

*João D. G. Azevedo [1,2], Rafael Ramos Uña [3], Ángela Barreda [3], Stuart Edwards [4], J. Iwan Davies [4], Jérôme Borme [1], Braulio Garcia-Cámara [3], Jana B. Nieder [1], Bruno Romeira [1,]**

[1] INL - International Iberian Nanotechnology Laboratory, Av. Mestre José Veiga, 4715-330, Braga, Portugal

[2] Department of Physics, University of Minho, Gualtar Campus, 4710-057, Braga, Portugal

[3] Group of Displays and Photonic Applications, Carlos III University of Madrid (UC3M), Av. de la Universidad, 30, 28911 Leganes, Spain

[4] IQE plc, Cardiff CF3 0LW, United Kingdom

## ABSTRACT

Next-generation on-chip photonic devices require compact, energy-efficient components integrating photodetection and light emission on silicon platforms. Achieving broadband photodetection combined with efficient emission in a single nanoscale device remains challenging. Here, we demonstrate elliptical nanopillar arrays monolithically integrated as III-V-on-Si templates, exhibiting broadband photoresponse across visible to near-infrared wavelengths, dynamic 'neuron-like' graded responses under modulated light, and pronounced electroluminescence at ~865 nm under forward bias. The nanostructured elliptical geometry approach enhances optical absorption and carrier dynamics within an active area significantly smaller than commercial detectors, enabling responsivities up to 0.55 A/W at 830 nm in compact nanoarrays. This work advances the development of miniaturized active nanophotonic devices suitable for integrated sensing, emitting and communication circuits, and neuromorphic processing on silicon substrates.

## Introduction

Nanostructured photodetectors and emitters are emerging as versatile platforms that exploit geometry, material anisotropy, and subwavelength sizes to achieve compact and spectrally-sensitive detection and efficient emission.[1] This will open new paths for ultracompact on-chip sensing, optical communications and neuromorphic or artificial-intelligence-based photonic systems.[2] However, the compact integration of nanoscale photodetectors and nano-emitting devices within on-chip electronics remains challenging. Key obstacles include managing of the lattice mismatch between III-V materials and silicon (Si),[3,4] achieving high electrical injection efficiency for light emission,[5] minimizing optical losses,[6] ensuring scalability,[1] reducing thermal noise and instability, and preventing unwanted crosstalk.[7] Additionally, current implementations of such active sensing-emitting units devices in computation platforms often lack the necessary nonlinearity and reconfigurability required for advanced functionalities.[8]

Several efforts are ongoing for the direct epitaxial growth or hybrid integration of nanoscale III-V and III-nitride structures, such as nanowires or nanoneedles, on a silicon substrate. Gallium arsenide (GaAs) and its ternary alloys, such as AlGaAs and InGaAs, have been grown on Si for devices such as nanowire light-emitting diodes (LEDs) in the near-infrared (NIR),[9] nanowire array solar cells and photodetectors,[10,11] and nanoneedle-based avalanche photodiodes[12]. Gallium nitride (GaN) has also been reported as an emerging material to implement on-Si-grown or transferred nanowires for visible light emission,[13] suitable for imaging applications, or photodetection in the visible and ultraviolet (UV) range for neuromorphic computing platforms.[14–16] Additionally, phosphorus and indium compounds (GaP/GaAsP/InGaP) have been analyzed in nanowires for tunable emission in the visible to NIR range and applications in solar cells, water splitting devices, and photodetectors.[17] Despite these advances, direct epitaxial growth of individual III–V/N nanostructures remains the dominant strategy. Yet, wafer-scale growth on Si, particularly of the III–V material, followed by top-down definition of nanoscale devices offers a potentially more scalable integration pathway. However, achieving wafer-scale epitaxial growth and nanoscale devices on a III-V-on-Si platform that simultaneously provides light emission and broadband photodetection, namely in the visible and near-infrared, remains challenging.

Light emission from nanoscale III-V and III-V-on-Si devices is particularly sensitive to non-radiative surface recombination at unpassivated surfaces,[18,19] and to device-to-device heterogeneity associated to the growth techniques and fabrication of nanostructures (e.g., nanoneedles or nanowires).[12] Nanoscale III-V and III-V-on-Si photodetectors have meanwhile been demonstrated across wavelength spanning from the visible to the infrared,[11,20] including communication between dual-function nanowires acting as both emitters and detectors.[16,20] Nevertheless, in contrast to bottom-up-grown nanostructures, the combined realization of broadband visible-to-NIR photodetection and NIR electroluminescence in top-down p-i-n nanopillar arrays monolithically integrated on Si has not yet been systematically investigated.

In this work, we report wafer-scale GaAs/AlGaAs n-i-p heterostructures grown on Ge-on-Si, exploring the lattice matching between Ge and III-V semiconductors, followed by nanofabrication of sub-micrometer elliptically shaped pillars. The nanoarray devices demonstrate highly efficient broadband photodetection from visible to near-infrared wavelengths, reaching a responsivity of 0.55 A/W at 830 nm. Under modulated light injection, the devices exhibit multi-pulse dynamics, exhibiting a cumulative increase in photocurrent, analogous to graded neuron-like responses,[21] of interest for nonlinear computing units to perform dynamic signal processing functions. Under forward bias, the same nanopillar arrays exhibit pronounced electroluminescence centered at ~865 nm with comparable emission as nanoarrays in native GaAs substrates, thereby combining broadband photodetection, dynamic optical response, and near-infrared emission within a single monolithically integrated III-V-on-Si nanophotonic platform. Since with elliptical nanopillars we are able to break in-plane symmetry at the unit-cell level, the geometrical anisotropy enables the potential for enhancing field confinement and light emission compared to planar films.[22–26] and provides a route toward polarization-dependent (dichroic) photodetection applications.[27]

## EPITAXIAL GROWTH

The III-V-on-Si hetero-epitaxial growth template was realized via a Ge-on-Si process performed on an ASM Epsilon chemical vapor deposition (CVD) tool at an industrial foundry (IQE plc). This process yielded a germanium surface with lattice match of GaAs on a 150 mm semi-insulating (SI) silicon substrate, with threading dislocation density (TDD) as low as $1\times10^{7}$ $cm^{-2}$. The epitaxial III-V stack for photodetection and emission (Figure 1a) consists of two main regions: a top n-i-n AlAs/GaAs/AlAs double barrier quantum well (DBQW), to investigate resonant-tunneling-induced nonlinear carrier transport[28] (beyond the scope of this work), and a bottom n-i-p GaAs/AlGaAs structure with a 280 nm thick bulk GaAs active layer for optical absorption and emission. The III-V stack growth was performed on an Aixtron G3 reactor by metal-organic chemical vapor deposition (MOCVD) also by IQE plc. In the same epitaxial growth run, the identical III-V heterostructure was also grown on native GaAs substrate, enabling a direct comparison with the III-V-on-Si platform.

## DESIGN AND FABRICATION

As shown in Figure 1b, we designed rectangular-lattice arrays of elliptical pillars with shortest diameter ($d_x$) between 150 nm and 500 nm, and longest diameter ($d_y$) between 300 nm and 2000 nm. The ratio between $d_x$ and $d_y$ is hereby referred to as 'axial ratio' and takes the values of 1:2 or 1:4. The distance between the centers of the pillars (pitch), was set between 2 and 5 times $d_x$ along the $x$ direction ($p_x$), and between 1.5 and 5 times $d_y$ along the $y$ direction ($p_y$). We chose these values to accommodate the dimensional limitations of our fabrication process while optimizing the arrangement of varying pillar sizes within uniform contact areas. The fabrication of the pillars consisted of a top-down approach employing e-beam lithography patterning and reactive ion etching (RIE), followed by the deposition of a 100 nm thick passivation silicon nitride ($Si_xN_y$) layer. Indium tin oxide (ITO) electrical contacts were patterned by ultra-violet (UV) lithography and deposited by physical vapor deposition on top of the pillars, together with gold (Au) contact pads. Extended details about the epilayer stack, mask layout and fabrication steps can be found in Section S1 of the Supporting Information. In Figure 1 we present a scanning electronic microscopy (SEM) images of a nanopillar array after the deposition of the ITO contact ($d_x$ = 300 nm and $d_y$ = 900 nm), panel c, and of a finalized device ($d_x$ = 200 nm and $d_y$ = 1000 nm), panel d, having bottom contacts pads in a titanium-platinum-gold (Pt/Ti/Pt/Au) multilayer and an additional gold pad on the top contact (not visible in Figure 1d, but partially covering the area beneath the tapered region as shown in Figure S1 of the Supporting Information). SEM images of the etched pillars reveal that the pitch factor influences the etching sidewall angles: arrays with $p_x = 2$ or $p_x = 3$ exhibit relatively steep sidewalls (slope < 10° from the pillar axis, case of Figure 1c), whereas with $p_x = 5$ the sidewall angle may reach ~20°.

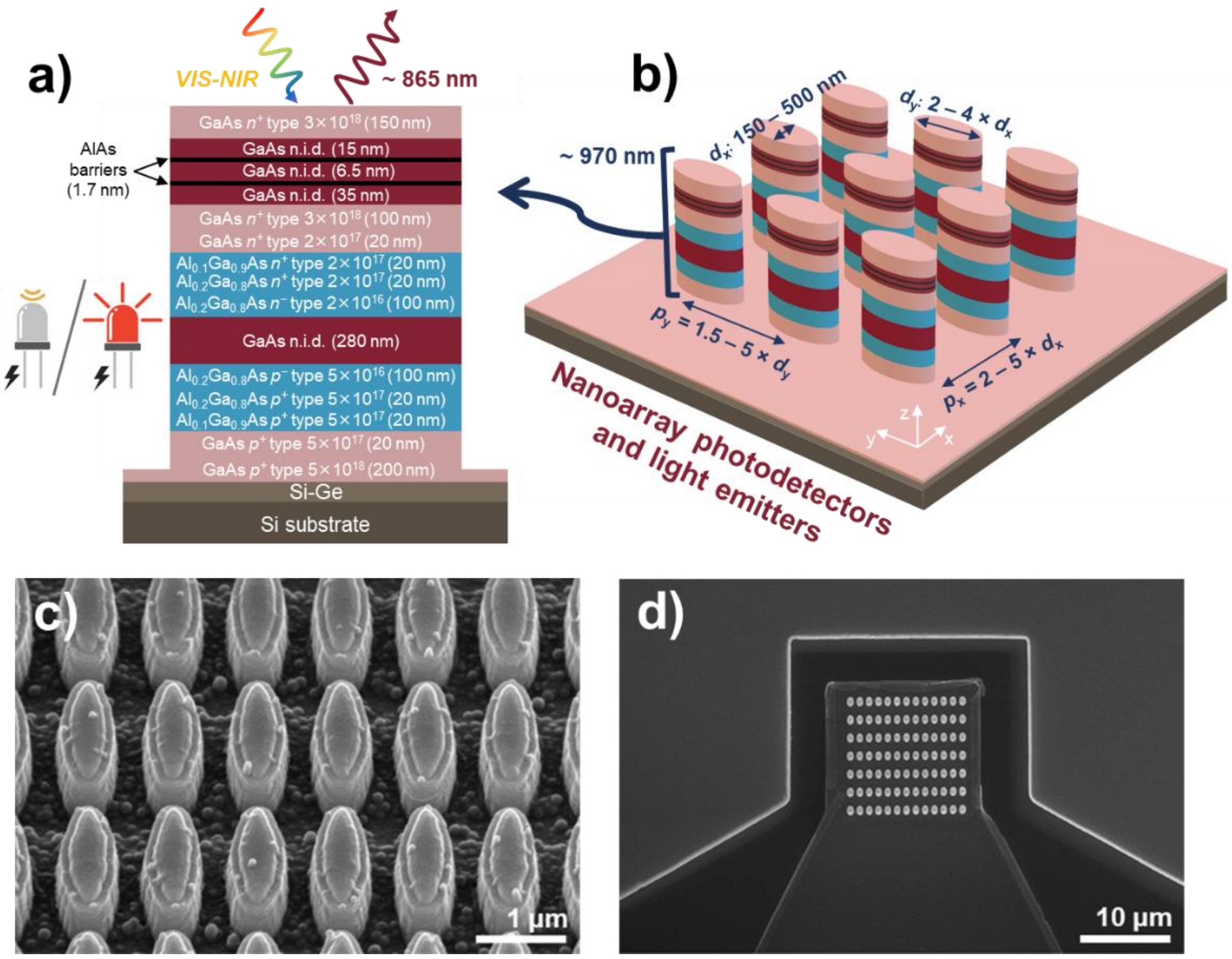

**Figure 1.** Design and fabrication of the III-V-on-Si elliptical nanopillar array devices. a) Full semiconductor layered structure employed in this study: semi-insulating Si substrate with lattice-matching Ge-on-Si platform, and n-i-p III-V epilayer stack. b) Schematic representation of the designed nanopillar arrays. Labels indicate the fabricated dimension ranges. c) Detailed scanning electron microscope (SEM) image of a nanopillar array ($d_x$ = 300 nm and $p_x$ = 3) after the deposition of the top ITO contact, taken with a tilting angle of 40° from the pillar axis. d) SEM top view image of a fully fabricated device, comprising the nanopillar array ($d_x$ = 200 nm and $p_x$ = 5), ITO top contact (inner pad) and Pt/Ti/Pt/Au ground contact (outer pad).

## RESULTS AND DISCUSSION

### Static Photoresponse

We comprehensively studied the light detection performance of our nanopillar arrays under continuous wave (CW) illumination using a manually positioned coplanar ground-signal-ground (G-S-G) electrical probe to drive the devices and a lensed optical fiber for light injection. Laser sources at 635 nm, 785 nm and 830 nm were used, with input spot diameters ranging from 2.5 µm to 25 µm (details about the setup components in Section S2). Figure 2a shows the measured photocurrent-voltage characteristic curves of two of the tested nanopillar arrays ($d_x = 150$ nm and $p_x = 2$; $d_x = 400$ nm and $p_x = 2$) under the same illumination conditions: injection at 635 nm with input power ($P_{in}$) in the 0.4 mW − 2.5 mW range. In these cases, and as for the remaining tested devices, characteristic 'S'-shaped traces are visible in reverse bias and before the turn-on voltage. The corresponding absolute and dark currents under reverse bias are presented in Figure S3a. While the photocurrent intensity is always proportional to the injected power, the inflexion point of the traces starts approaching the turn-on voltage (shifts to the right) as pillar size increases. In Figure 2b, for the same arrays, we plot the measured absolute photocurrent as a function of the effective input optical power ($P_{eff}$) in reverse voltage bias ($V_{bias}$), from 0 V to −4 V. $P_{eff}$ represents the optical power effectively incident on the active semiconductor region of the pillars, as defined by Equation (S1) of the Supporting Information. A quasi-linear trend is observed for the displayed injection power range ($P_{eff} < 0.5$ mW), while photocurrent saturation becomes apparent at higher injection power levels (see additional results of Figure S3b). We attribute this saturation to the continuous accumulation of charge, which prevents efficient collection of new photogenerated carriers, an effect known as 'carrier screening'.[29]

In Figure 2c, we plot the absolute photocurrent obtained with a representative nanopillar array ($d_x$ = 400 nm and $p_x$ = 2) for the three tested wavelengths. Similar monotonous responses were observed, showing effective photodetection operation at the three wavelengths tested in this work. We compare the photoresponse of different nanopillar array configurations by quantifying their responsivity, $R$, a standard metric of their detection efficiency (defined by Equation (S2)). In Figure 2d, we plot the responsivity values for the tested nanopillar arrays at 0 V and −4 V bias voltages (hollow and solid disk markers, respectively), under illumination at (i) $\lambda$ = 635 nm, (ii) $\lambda$ = 785 nm, and (iii) $\lambda$ = 830 nm, as a function of $d_x$ and $p_x$. In general, $R$ increases with both $d_x$ and bias voltage, a trend that was equally observed for pillars with 1:2 and 1:4 axial length ratios (points color-coded according to both gradients in the figure).

In Table S2 of the Supporting Information, we benchmarked the most photosensitive arrays in III-V-on-Si ($d_x$ = 500 nm and $p_x$ = 5; $d_x$ = 400 nm and $p_x$ = 2, highest points in Figure 2d(i-iii)) against previously reported nanofabricated devices. Our highest responsivity nanopillar arrays reach 0.55 A/W in the 800 nm – 900 nm range at 4 V, comparable to the 0.57 A/W – 0.75 A/W reported for GaAs/AlGaAs nanowire and nanoneedle photodetectors operated at 5V – 10 V. Notably, the compact nanopillar arrays exhibit responsivity values comparable to commercially available GaAs photodetectors (detailed in Table S3), thus offering a substantially reduced device footprint. For comparison, nanopillar arrays fabricated in parallel on native GaAs substrates using the same epilayer stack exhibited 1.3- to 2-fold higher photoresponsivities than the III–V-on-Si nanoarrays (Figure S4). This potentially reflects differences in crystalline quality, although fabrication-induced variability cannot be excluded, and indicates further potential for improvement of the photodetection in the III-V-on-Si platform.

We also compared $R$ in arrays whose pillars have the same $d_x$ (400 nm) but different $p_x$ (2 vs 5) and axial ratio (1:4 vs 1:2). The respective points, connected in Figure 2d(i) by gray dotted lines, reveal higher $R$ for the most compact layout ($p_x = 2$ and axial ratio of 1:4). This observation was validated by a numerical simulation study and additional photoluminescence (PL) experiments, whose results are combined in Figures S5a-c (detailed discussion in Section S4). A general trend is detected: for the same $d_x$, a lower pitch configuration leads to higher absorption, which translates into higher $R$. Additionally, a more elongated nanostructure can provide a longer effective optical cross-section, increasing the in-coupling efficiency of incident light and enhancing carrier photogeneration.

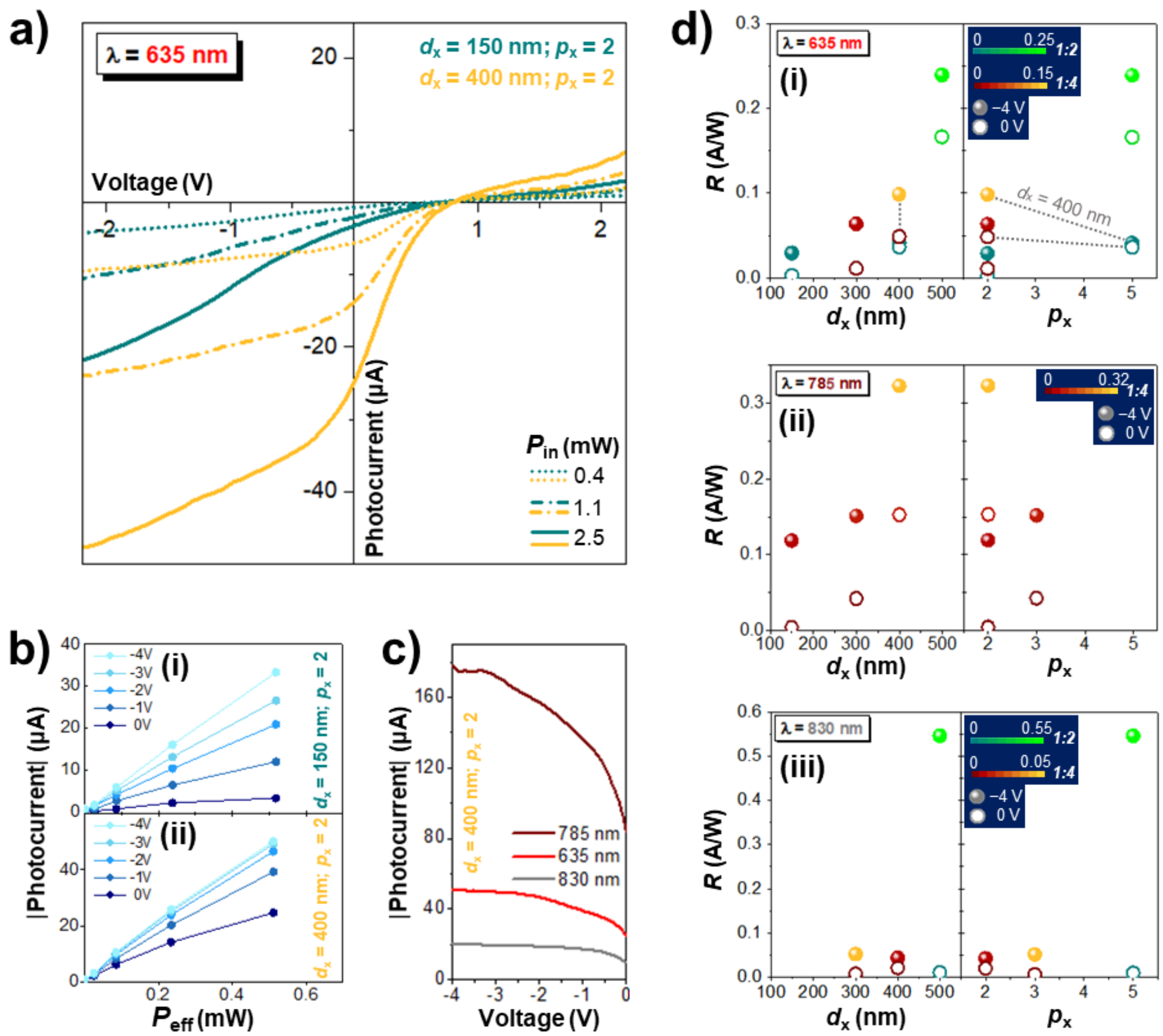

**Figure 2.** Static photodetection characteristics of the tested III-V-on-Si elliptical nanopillar arrays under continuous wave illumination. a) Photocurrent-voltage traces obtained at 635 nm (using light input powers of 0.4 mW, 1.1 mW and 2.5 mW) for different nanopillar arrays ($d_x$ = 150 nm and $p_x$ = 2; $d_x$ = 400 nm and $p_x$ = 2). b) Variation of the absolute photocurrent with the effective injected optical power ($P_{eff}$) of the two arrays mentioned in a), respectively (i) and (ii). c) Absolute photocurrent across the tested reverse bias voltage range for the nanopillar array having $d_x$ = 400 nm and $p_x$ = 2, illuminated at 635 nm, 785 nm and 830 nm ($P_{in}$ = 2.5 mW). d) Calculated responsivity values ($R$) for each tested nanopillar array illuminated at (i) 635 nm, (ii) 785 nm, and (iii) 830 nm ($P_{in}$ = 2.5 mW), with $V_{bias}$ = 0V (hollow disks) and $V_{bias}$ = −4V (solid disks) – Left: $R$ vs $d_x$; Right: $R$ vs $p_x$. Gray dotted lines connect results obtained with same $d_x$ but different $p_x$.

## Dynamic Photoresponse

Following this analysis, we evaluated the dynamic photoresponse of the nanopillar arrays using intensity modulated optical input signals at 830 nm. Figure 3a depicts the implemented experimental setup, comprising direct modulation of the laser light input and separation of the DC and RF signals using a bias-T. First, we studied the photogenerated voltage response to single pulses. For this test, the laser was directly modulated by a 100 ns square-wave electrical pulsed signal. An example of the single pulse input signal used to drive the laser is plotted in Figure 3b(i), using peak-to-peak voltage ($V_{pp}$) of 5 V and $P_{in}$ = 2.5 mW. In Figures 3b(ii) and 3b(iii), we show responses to single pulse inputs for the array having $d_x$ = 150 nm and $p_x$ = 2 (using $V_{pp}$ = 5 V) and for the array having $d_x$ = 300 nm and $p_x$ = 3 (using $V_{pp}$ = 2 V), respectively, at −4 V bias. The voltage profiles of the devices exhibit quasi-linear ramps throughout the pulse duration, followed by single exponential decay curves that were fitted to estimate the fall times.

In Figure 3c, we plot the fall times calculated from a monoexponential fit of the decay curves obtained with several tested nanopillar arrays, across $d_x$ and $p_x$. The calculated values fall within the 150 ns − 430 ns range, slightly shorter than those reported in literature on GaAs-based micropillars comprising an n-i-n stack with an AlAs/GaAs/AlAs active region (250 ns − 660 ns).[30] We observe a general increase of the fall times with $d_x$, consistent with the scaling of lateral diffusion paths with pillar size.[31] Once again, we evaluated results obtained in arrays with the same $d_x$ (300 nm) but different $p_x$ (2 vs 3), applying $V_{pp}$ = 2 V. In Figure 3c, the respective points are connected by a gray dotted line, and reveal a longer fall time for the most compact array ($p_x$ = 2).

Next, and as shown in Figure 3d, we explored the arrays' rise and decay dynamics by testing a four-pulse input. Using a pulse duration of 20 ns and an inter-pulse delay of 50 ns (represented in Figure 3d(i)), we present the measured output voltage obtained with arrays having $d_x$ = 300 nm and $p_x$ = 2 (Figure 3d(ii), using $P_{in}$ = 2.5 V, $V_{pp}$ = 2 V, $V_{bias}$ = −4 V), and $d_x$ = 250 nm and $p_x$ = 2 (Figure 3d(iii), using $P_{in}$ = 6.1 V, $V_{pp}$ = 5 V, $V_{bias}$ = −6 V). In the first case, the observed output profile is a cumulative series of voltage steps that reaches ~70% of the measured input amplitude, followed by an expected exponential decay after the final pulse. In the second case, however, the response exhibits a sawtooth-like wave shape, with a rapid saturation at ~40% of the input amplitude and a visible decay below 0 V before stabilization.

Consequently, we quantified the voltage steps ($\Delta V$) measured between consecutive pulses across several experimental sets. In Figure 3e, each quadrant depicts results obtained under different tested experimental conditions, namely peak-to-peak amplitude, injected optical power of the input signal and bias voltage of the device. Regardless of the conditions, we observed that $\Delta V$ decreases continuously for every tested array. Using $P_{in}$ = 2.5 mW, $V_{pp}$ = 2 V and $V_{bias}$ = −4 V (case of Figure 3e(i)), the measured voltage steps remain above 0 V throughout the pulse sequence, indicating a non-saturating charge-accumulation function. In opposition, with $P_{in}$ = 6.1 mW, $V_{pp}$ = 5 V and $V_{bias}$ = −6 V (Figure 3e(iv)), the voltage actually drops below 0 V in the last step for the two tested arrays, framing an over-driven limit scenario that suppresses the integrator response.

In general, the extracted fall times of the signals presented in Figures 3b and 3d are two orders of magnitude longer than previously reported electroluminescence results for GaAs micro- and nanopillar devices (from hundreds of picoseconds to nanoseconds).[19,30] We associate long decay to charge generation in the highly doped GaAs contact region, contribution to slow diffusion of photogenerated carriers,[32] In addition, accumulation of photo-generated charges (particularly the holes) at the barriers of the double barrier quantum well can prolong charge storage and has been reported to limit photodetectors response speed.[33] The *RC* time constant of the circuit has been estimated < 1 ns, based on epilayer stack materials, device geometry and respective electrical contacts (see Equation (S3)). Therefore, the transient response of the devices seems to be governed mainly by the carrier dynamics mentioned above and not limited by the extrinsic *RC* circuit parasitics.

The observed multi-pulse dynamics, in particular the cases where the voltage signal cumulatively increases (see Figure 3d(ii) as an example), reflect a carrier population build-up, due to fall times significantly exceeding the pulse interval. Said behavior could be potentially exploited for signal processing to mimic the leaky integration of a biological neuron in the context of graded neuron-like neuromorphic applications,[21] while also demonstrating the devices' capability to process time and frequency-based patterns of information. $\Delta V$ can be regarded as an indicator of the rate at which the generated carriers populate available energy states. Therefore, the quasi-linear slopes of Figure 3e suggest that the "leaky reservoirs" materialized by the nanopillar arrays are being filled at constant rates. With $V_{pp}$ = 5 V (Figures 3e(ii-iv)), $\Delta V$ is already 0 V by the fourth pulse, revealing a balance between carrier photogeneration and recombination during the inter-pulse time delay. In the limit scenario of Figure 3e(iv), the combination of high voltage bias and high input power leads to charge field screening near the doped interfaces and amplified Joule heating within the array.[34] This further reduces the carrier mobility and ultimately suppresses the device's ability to accumulate charge, resulting in saturation of the voltage signal and detector nonlinearity.

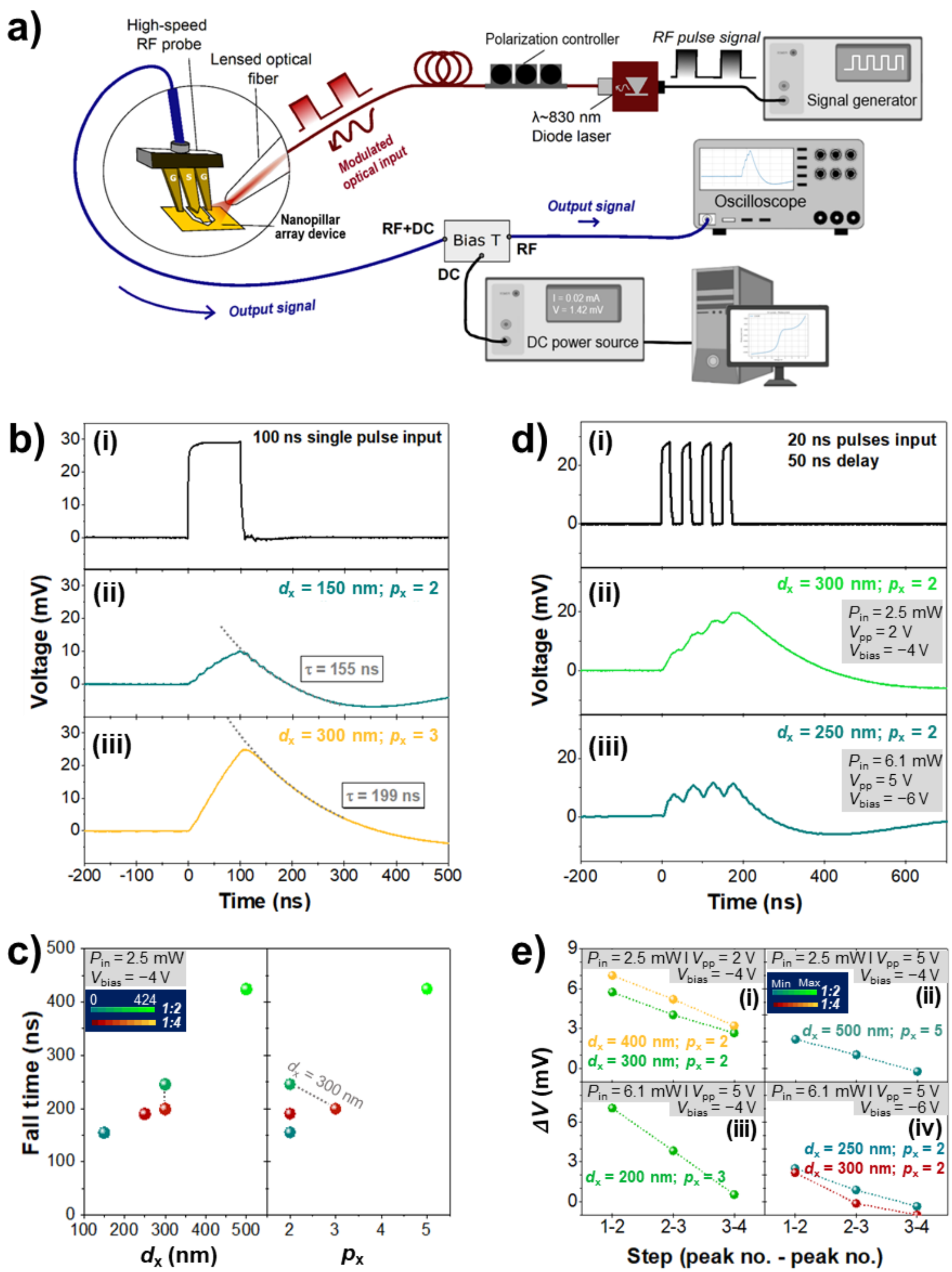

**Figure 3.** Dynamic photoresponse characteristics of the tested III-V-on-Si elliptical nanopillar arrays at 830 nm. a) Experimental setup used to assess the dynamic photoresponse of nanopillar array devices. b) Voltage traces obtained with a single 100 ns square pulse ($P_{in}$ = 2.5 mW): (i)

input signal ($V_{pp}$ = 5 V); (ii) response of a nanopillar array having $d_x$ = 150 nm and $p_x$ = 2 ($V_{pp}$ = 5 V; $V_{bias}$ = −4 V); (iii) response of a nanopillar array having $d_x$ = 300 nm and $p_x$ = 3 ($V_{pp}$ = 5 V; $V_{bias}$ = −4 V). Gray dotted traces indicate the fitting decay curves and insets show the respective fall times. c) Determined fall times for each tested nanopillar array ($P_{in}$ = 2.5 mW and $V_{bias}$ = −4 V) against $d_x$ and $p_x$. The results for $d_x$ = 300 nm (indicated by the gray dotted lines) were obtained with $V_{pp}$ = 2 V while the remaining ones were obtained with $V_{pp}$ = 5 V. d) Voltage traces obtained with a train of four square pulses: (i) input signal of 20 ns pulses and 50 ns inter-pulse time delay; (ii) response of a nanopillar array having $d_x$ = 300 nm and $p_x$ = 2 ($P_{in}$ = 2.5 V, $V_{pp}$ = 2 V, $V_{bias}$ = −4 V); (iii) response of a nanopillar array having $d_x$ = 250 nm and $p_x$ = 2 ($P_{in}$ = 6.1 V, $V_{pp}$ = 5 V, $V_{bias}$ = −6 V). e) Determined voltage steps of several nanopillar arrays as response to four square pulses, under different test conditions: (i) $P_{in}$ = 2.5 mW, $V_{pp}$ = 2 V, $V_{bias}$ = −4 V; (ii) $P_{in}$ = 2.5 mW, $V_{pp}$ = 5 V, $V_{bias}$ = −4 V; (iii) $P_{in}$ = 6.1 mW, $V_{pp}$ = 5 V, $V_{bias}$ = −4 V; (iv) $P_{in}$ = 6.1 mW, $V_{pp}$ = 5 V, $V_{bias}$ = −6 V. All traces were obtained with 20 ns pulses and 50 ns inter-pulse time delay, except for the case having $d_x$ = 400 nm and $p_x$ = 2 in (i), that was obtained with 50 ns pulses and 70 ns inter-pulse time delay.

**Electroluminescence**

Using the static photodetection setup, we also characterized light emission from forward biased nanopillar arrays, by using the lensed optical fiber to extract the emitted light. Figure 4a presents the emission spectra of one of the tested nanopillar arrays ($d_x$ = 150 nm and $p_x$ = 2), across a range of applied voltages (3.5 V – 5 V). The emission peak at 862 nm matches the room-temperature band-edge transition of the bulk emitter GaAs in the active layer ($E_g \approx 1.43$ eV). Additionally, it is in line with the peak emission obtained on nanopillar arrays fabricated on GaAs native substrate (867 nm), as plotted in Figure S6a, indicating sufficient epitaxial growth quality on the Ge-on-Si platform. The full width at half maximum (FWHM) of the measured peak reaches ~40 nm at 5 V, comparable with existing literature for the applied current density values (20 nm − 50 nm).[19] We plot the corresponding light-current-voltage (*L-I-V*) traces in Figure 4b, reflecting a characteristic diode-like behavior with currents in the orders of tens of μA throughout the voltage sweep and turn-on voltage around 3 V. The counterpart results on native GaAs substrate reveal an identical *L–I* progression and a clearly lower turn-on voltage (averaging around 1.5 V), as exemplified by the case presented in Figure S6b.

Across the tested array designs in III-V-on-Si, we observe that the turn-on voltage decreases to roughly 2 V when $d_x$ increases (see additional *L-I-V* traces of Figures S7a-b). This voltage span is an effect of each device's series resistance, which decreases as the top-face contact area of the pillars increases. The initial low-slope region of the *L–I* curve in Figure 4b indicates a gradual evolution in device efficiency due to initial non-radiative surface effects, which prevail before bimolecular radiative recombination becomes dominant.[35]

In Figure 4c, we plot an example of voltage-dependent spectra for an array with $d_x$ = 300 nm and $p_x$ = 5. The pronounced spectral asymmetry matches previously reported results for GaAs nanopillar arrays of comparable dimensions.[19] We observe a red-shift in peak emission (from 862 nm to 867 nm) related with carrier injection in the bulk emitter and thermal effects at higher injection conditions, and also potential non-negligible carrier recombination in the surrounding p+ type doped bottom layer (see respective photoluminescence spectrum in Figure S8). The current-to-light conversion efficiency of the nanopillar arrays was evaluated by normalizing the emitted intensity (detection counts) by the active area of each device and plotting it against the driving current density (traces of Figure S7c). After stabilization of light emission, the obtained electroluminescence (EL) intensity per unit area ($EL_A$, in counts $s^{-1}$ $\mu m^{-2}$) scales with the square of the current density according to the quadratic radiative recombination regime defined by Equation (1):

$$\mathrm{EL_A} = \alpha J^2 \quad (1)$$

Here, $\alpha$ (in counts $s^{-1}$ $\mu m^{-2}$ $\mu A^{-2}$) is a multiplication factor that quantifies the limit differential efficiency of the nanopillar arrays, while $J$ is the current density. In Figure 4d, we present the values for $\alpha$ of several tested arrays, across pillar size and array pitch, determined by fitting the limit slope of their *L–I* traces. While $p_x$ does not seem to significantly affect differential efficiency, we observe that only pillars with $d_x$ in the 200 nm – 400 nm range exhibit $\alpha$ above $10^{-1}$ counts $s^{-1}$ $\mu m^{-2}$ $\mu A^{-2}$, with the maximum value of 2 counts $s^{-1}$ $\mu m^{-2}$ $\mu A^{-2}$ being reached with $d_x$ = 200 nm and $p_x$ = 2. Such enhancement is potentially caused by the emission directionality of these specific array configurations, leading to an increase in fiber coupling efficiency. Moreover, optical confinement and the presence of efficient mode propagation can also be accounted for when the nanopillar dimensions become comparable to the effective wavelength in GaAs ($\lambda_{em}/n_{GaAs} \approx 245$ nm).[36]

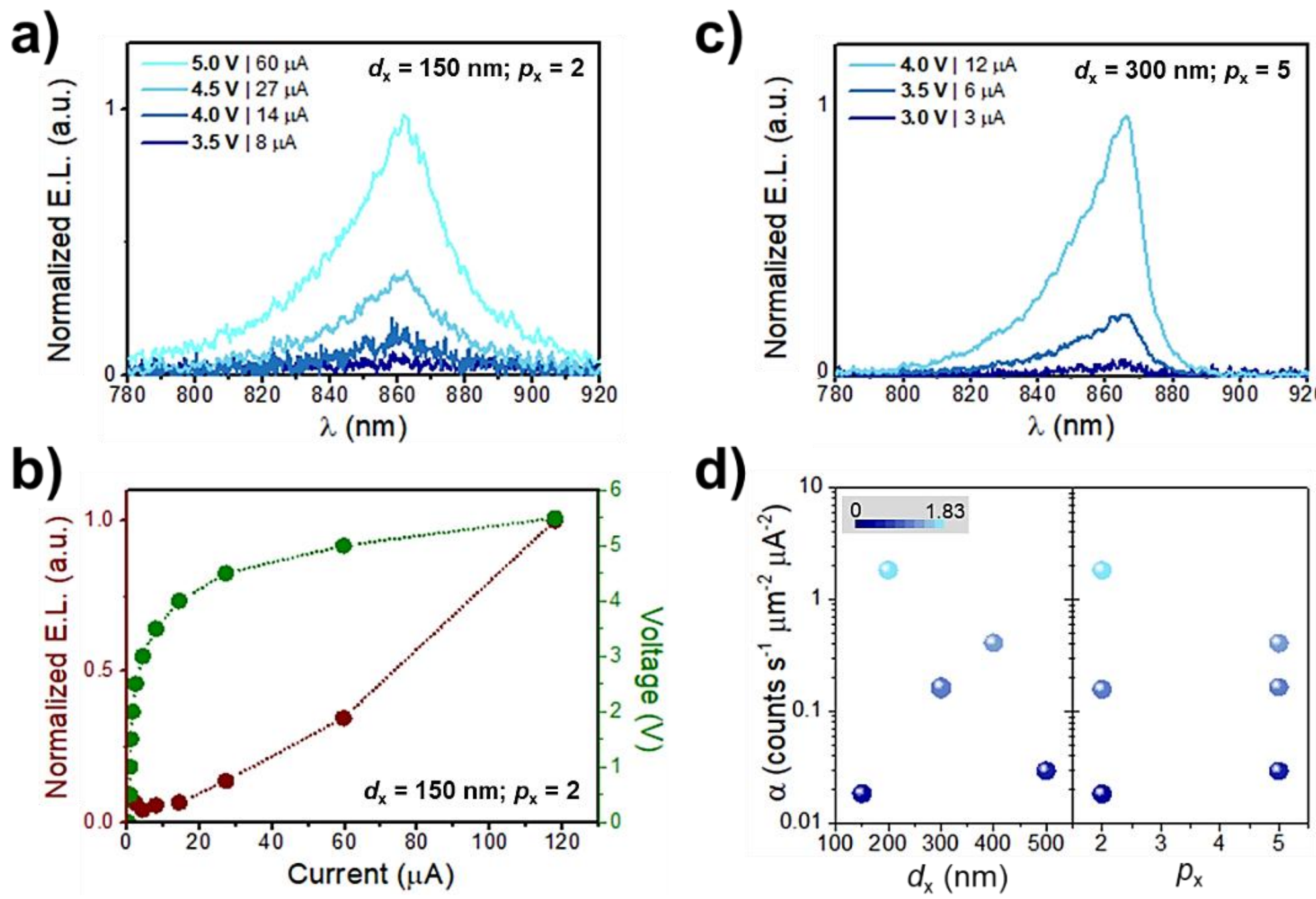

**Figure 4.** Static emission characteristics of the tested III-V-on-Si elliptical nanopillar arrays. a) Electroluminescence (EL) emission spectra of an array having $d_x$ = 150 nm and $p_x$ = 2 (1:4 axial ratio), obtained with 3.5 V ≤ $V_{bias}$ ≤ 5 V. b) Light-current-voltage (*L-I-V*) characteristics of the device mentioned in a). c) EL emission spectra of an array having $d_x$ = 300 nm and $p_x$ = 5 (1:2 axial ratio), obtained with 3 V ≤ $V_{bias}$ ≤ 5 V. d) Calculated differential efficiency coefficient ($\alpha$) for the tested nanopillar arrays across $d_x$ and $p_x$.

## CONCLUSION

In summary, we demonstrate the design, fabrication and characterization of top-down elliptical n-i-p GaAs/AlGaAs nanopillar arrays monolithically integrated on a Ge-on-Si platform for dual photodetection and light emission. At zero and low reverse bias, the devices exhibit significant photocurrent from visible to near-infrared wavelengths, with responsivities of 0.24 A/W at 630 nm, 0.32 A/W at 785 nm, and 0.55 A/W at 830 nm. Under modulated light, the nanopillar arrays exhibit sub-microsecond fall times in the range of 150 ns – 430 ns where we have shown cumulative photocurrent dynamics of interest for dynamic signal processing. Under forward bias, we report light emission at ~865 nm, with identical emission properties of designs on native GaAs. These results establish a compact III–V-on-Si nanophotonic platform combining broadband photodetection, dynamic optical response, and near-infrared emission, with potential for integrated sensing, optical communication, and neuromorphic photonic circuits, addressing the current demands for miniaturization and energy efficiency.

ASSOCIATED CONTENT

**Supporting Information.**

Detailed description of the epilayer stack configuration and nanopillar array fabrication process, including mask design; description of the used setups for light detection and emission studies; additional analysis of light detection characteristics on selected nanopillar arrays; Finite-Difference Time-Domain (FDTD) simulation of photoabsorption in selected nanopillar arrays; additional photoluminescence (PL) and electroluminescence (EL) results (PDF)

AUTHOR INFORMATION

**Corresponding Author**

* Correspondence: bruno.romeira@inl.int

## Author Contributions

The study was conceptualized and conceived by J.D.G.A. and B.R. R.R.U. conducted the FDTD photodetection simulation study in collaboration with J.D.G.A. S.E. and J.I.D. provided the raw III-V epilayer stacks on GaAs and on Si substrate using Ge-on-Si. J.D.G.A designed and fabricated the elliptical nanopillar devices in cleanroom environment, and carried out static and dynamic photoresponse studies, light emission studies, and data analysis, in feedback with B.R.. J.B. carried out e-beam exposure of the samples. B.R. provided fundamental understanding of III-V material dynamics and experimental guidance. A.B. and B.G.-C. contributed to data analysis and interpretation of the results. B.R. and J.B.N. coordinated project administration and funding acquisition and allocation. The manuscript was written by J.D.G.A and all authors contributed to its final version.

## Funding Sources

The authors received support for this project from HORIZON EUROPE – InsectNeuroNano; Portuguese Foundation for Science and Technology (FCT) – META-LED and ERC-PT-A; Spanish State Research Agency (AEI); MICINN – Ramon y Cajal Fellowship.

## Notes

The authors declare no competing financial interest.

ACKNOWLEDGMENT

The authors acknowledge access and support from the Micro and Nanofabrication Facility and the Nanophotonics and Bioimaging Facility at INL. J.D.G.A. acknowledges the support in frame of the PhD program in Physics at the University of Minho. R.R.U. wants to express his thanks for the Spanish National FPI grant. This work was funded by the European Union, HORIZON EUROPE Framework Programme, Project 101046790 – InsectNeuroNano, by the Portuguese Foundation for Science and Technology (FCT), projects 2022.03392.PTDC – META-LED and ERC-PT-A 13/C06-i06/2024.P1295330416, and by the Spanish State Research Agency (AEI), project no. PID2022-137857NA-I00. R.R.U. thanks for the FPI grant linked to Spanish national project (PID2022-137857NA-I00) A.B. also thanks MICINN for the Ramon y Cajal Fellowship (grant No. RYC2021-030880-I).